\documentclass[aps,prl,reprint,amssymb,nofootinbib]{revtex4-2}
\usepackage[T1]{fontenc}
\usepackage{amsmath}
\usepackage{graphicx}
\usepackage{xcolor}

\begin{document}


\title{Covariant virtual work and the d’Alembert–Lagrange formulation of general relativity}

\author{A.~Cuevas}
\email{angel.cuevas@fisica.uaz.edu.mx}

\author{C.~Ortiz}

\affiliation{
Unidad Acad\'emica de F\'isica,
Universidad Aut\'onoma de Zacatecas,
Solidaridad esq.\ la Bufa S/N,
Zacatecas, Zac., M\'exico, 98060
}

\date{\today}

\begin{abstract}
We develop a covariant virtual work structure and a corresponding
d’Alembert–Lagrange principle for spacetime geometry. Within this framework, General Relativity arises as a particular realization
of the principle, leading to a covariant d’Alembert–Lagrange formulation in which the
Einstein field equations arise from the vanishing of total covariant
virtual work on admissible metric variations, rather than from action
extremality.
The covariant virtual work structure provides a covariant classification of constraint-induced
contributions, distinguishing ideal reactions, which perform no virtual
work, from non-ideal sectors contribute explicitly to it. The structure extends naturally to one-sided admissibility conditions, yielding a covariant inequality structure.
Constraints generate reaction terms. In
particular, an isoperimetric constraint produces a cosmological term as
an ideal reaction fixed by spacetime averages, so that the cosmological
constant emerges as a global parameter determined by admissibility,
reflecting an intrinsically nonlocal geometric origin.
\end{abstract}
\maketitle

\textit{Introduction.—} General Relativity (GR) was originally derived by Albert Einstein heuristically \cite{Einstein1915}. David Hilbert subsequently formulated GR using Hamilton’s principle, from which the Einstein field equations (EFE) follow by extremizing the Einstein–Hilbert action under arbitrary metric variations with suitable boundary conditions \cite{Hilbert1915,wald1984general}.

However, the Hamilton formulation is structurally limited. In its standard form, the stationary action principle does not account for dissipative effects, which must be introduced ad hoc, requires integrability, and does not uniquely determine the boundary terms of the action. In GR, this leads to a non-unique specification of boundary contributions, which must be fixed by adding boundary terms to render the variational problem well posed \cite{GibbonsHawking1977,Chakraborty:2016yna,Parattu:2015gga,Parattu:2016trq}.

Extensions to Hamilton’s principle incorporate constraints through Lagrange multipliers \cite{Papastavridis2002,goldstein2002classical}. However, for nonintegrable constraints, constrained extremization generally does not reproduce the physically realized dynamics \cite{Bloch2015,Zampieri1998,Flannery2011}. Consequently, it does not provide a fully general framework for constrained variational dynamics; this broader role belongs instead to analytical mechanics.

Analytical mechanics provides the general area of study in which variational principles are formulated and organized \cite{lanczos1970variational,Gantmakher1970,Cline,Papastavridis2002}. The principle of virtual work defines the fundamental structure \cite{Bernoulli1717}, from which the d’Alembert–Lagrange principle  follows as its extension to dynamics \cite{dAlembert1743,Lagrange1788}, expressing the equations of motion as an mechanical equilibrium condition in configuration space through the vanishing of total virtual work over admissible variations. This formulation does not presuppose integrability and naturally incorporates constraint forces and non-conservative effects, while recovering Hamilton’s principle as a particular case under suitable conditions \cite{lanczos1970variational}. 

In modern geometric language,  virtual work structure is formulated through the pairing between tangent and cotangent spaces of configuration space \cite{AbrahamMarsdenRatiu1988,BulloLewis2005,Romano2009}. Admissible variations define a subspace of the tangent space, while generalized forces act as covectors on these variations and belong to the annihilator of the admissible subspace \cite{Bloch2015,Zubelevich2020,LeokOhsawa2011,Yoshimura2006DiracSI}. Mechanical equilibrium is thus characterized by the vanishing of this pairing. In classical settings, this structure is often represented through a metric-induced inner product on configuration space \cite{arnold1989mathematical}.

In this \textit{Letter} we develop a covariant d’Alembert–Lagrange principle for spacetime geometry, formulated in terms of a covariant virtual work structure together with a mechanical admissibility criterion encoded in Postulate~A. Within this framework, GR is recovered as a particular realization of the principle, in which the EFE emerge as local equilibrium conditions in configuration space from the vanishing of total covariant virtual work, rather than from an extremality condition imposed on the action. This formulation extends beyond the stationary action principle, allowing for structures not naturally captured within standard variational approaches. The formalism further admits one-sided admissibility conditions, under which the set of admissible variations is restricted and the total covariant virtual work satisfies an inequality condition. This structure yields a covariant analogue of Fourier’s inequality \cite{lanczos1970variational,Chetayev1989}. 

As an application of the framework to GR, a global admissibility constraint is imposed through an isoperimetric condition \cite{CourantHilbert1953,GelfandFomin1963}, from which a cosmological term arises as an ideal reaction associated with this integral constraint, whose value is fixed by a spacetime average of the trace sector, thereby acquiring a global mechanical origin.

\textit{Covariant virtual work.—} A covariant virtual work structure requires a precise geometric identification of configuration space and of the admissible virtual displacements defined upon it.

We consider as configuration space the space of smooth Lorentzian
metrics on a fixed differentiable manifold $M$,
\begin{equation}
\mathcal C := \mathrm{Met}(M).
\end{equation}
An element $g_{\mu\nu}\in\mathcal C$ represents a geometric
configuration of spacetime. At this stage no dynamical condition is
imposed; in particular, configurations are not assumed to satisfy any
field equations (off--shell\footnote{We follow the standard usage in the literature, where configurations not satisfying the equations of motion are termed \emph{off-shell}, while those satisfying them are termed \emph{on-shell}; see, e.g., \cite{Khavkine_2013,Barnich:2020ciy}.}).

A general metric variation defines a tangent vector in configuration
space,
\begin{equation}
\delta g^{\mu\nu} \in T_g\mathcal C .
\end{equation}

When constraints are imposed, they restrict the admissible configurations.
We denote by
\begin{equation}
\mathcal C_{\mathrm{adm}} \subset \mathcal C,
\end{equation}
the subset satisfying these conditions. Admissible virtual displacements
are metric variations that preserve the constraints to first order.
Accordingly, any variation decomposes as
\begin{equation}
\delta g^{\mu\nu}
=
\delta g^{\mu\nu}_{\mathrm{adm}}
+
\delta g^{\mu\nu}_{\perp},
\end{equation}
where $\delta g^{\mu\nu}_{\mathrm{adm}}$ is compatible with the constraints
and $\delta g^{\mu\nu}_{\perp}$ violates them, reflecting the standard decomposition in constrained dynamics \cite{Djukic2012}.\footnote{A covariant parametrization is
$\delta g_{\mathrm{adm}}=\mathcal L_X g + h$, with $h\in\mathcal H_g$
spanning complementary admissible symmetric directions \cite{York:1973ia,Lee:1990nz}. No quotient by
diffeomorphisms is performed.}

Given a symmetric tensor $A_{\mu\nu}$, interpreted as a generalized
covariant force density on configuration space, we define the local covariant
virtual work density as
\begin{equation}
\delta w[A]
\equiv
A_{\mu\nu}\delta g^{\mu\nu}_{\mathrm{adm}},
\label{eq:TV-local}
\end{equation}
and the covariant virtual work on a region $V$ as
\begin{equation}
\delta W[A]
=
\int_V d^4x\,\sqrt{-g}\,A_{\mu\nu}\delta g^{\mu\nu}_{\mathrm{adm}} .
\label{eq:TV-total}
\end{equation}

Admissible virtual displacements are not physical perturbations nor
dynamical evolutions, but infinitesimal probes of admissible directions
in configuration space used to test mechanical equilibrium.

A configuration $g_{\mu\nu}$ is in mechanical equilibrium if
\begin{equation}
\delta W[A]=0,
\qquad
\forall\,\delta g^{\mu\nu}_{\mathrm{adm}} .
\end{equation}
This condition defines the covariant principle of virtual work for admissible variations.

For compactly supported admissible variations in the interior of $V$,
this condition probes all local admissible directions and expresses the
vanishing of the pairing between $A_{\mu\nu}$ and admissible
variations \cite{Bloch2015,LeokOhsawa2011},
\begin{equation}
\langle A , \delta g_{\mathrm{adm}} \rangle = 0,
\qquad
\forall\, \delta g_{\mathrm{adm}} \in T_{\mathrm{adm}}\mathcal C .
\end{equation}
Here $A\in T_g^*\mathcal C$ and $\delta g_{\mathrm{adm}}\in T_g\mathcal C$.

In analytical mechanics, Postulate~A states that reaction forces associated
with \emph{ideal} constraints perform no virtual work on admissible
displacements \cite{Gantmakher1970,lanczos1970variational,Papastavridis2002,FassoSansonetto2010}.
This condition characterizes ideality rather than constituting a universal
property of all constraint reactions.

Within the covariant formulation, Postulate~A requires that ideal reaction
tensors perform no virtual work on admissible virtual displacements,
\begin{equation}
\delta W_{\mathrm{reac}}
=
\int_V d^4x\,\sqrt{-g}\,
C_{\mu\nu}\,\delta g^{\mu\nu}_{\mathrm{adm}}
=0 ,
\label{eq:PostulateA}
\end{equation}
for all admissible variations $\delta g^{\mu\nu}_{\mathrm{adm}}$.

This provides a covariant criterion of ideality: such tensors lie in the
annihilator of the admissible variation space. More generally, reaction
tensors may be non-ideal, characterized by
\begin{equation}
\delta W_{\mathrm{reac}} \neq 0,
\end{equation}
while preserving the kinematical definition of admissible variations.

This viewpoint is consistent with classical treatments in which non-ideal
constraints do not invalidate the d’Alembert–Lagrange framework but require
its extension beyond the ideality assumption
\cite{Gantmakher1970,Papastavridis2002}. 
In modern approaches, non-ideal effects enter at the level of the equations
of motion \cite{UdwadiaKalaba2000,Udwadia2002,Udwadia2005,SolominZuccalli2005};
here they are characterized directly at the level of the covariant
virtual work.

The covariant virtual work principle does not presuppose any specific gravitational dynamics.
For a diffeomorphism-invariant Lagrangian density $\mathcal L$, one may define
\begin{equation}
A_{\mu\nu}
:=
-\frac{2}{\sqrt{-g}}
\frac{\delta S}{\delta g^{\mu\nu}},
\qquad
S=\int_V d^4x\,\sqrt{-g}\,\mathcal L.
\end{equation}
Lovelock’s theorem constrains the class of Lagrangians yielding
second-order field equations \cite{Lovelock1971}. This restriction acts at the
\emph{Lagrangian level}, whereas the present  principle operates at the
\emph{mechanical level}, where equilibrium is characterized by the vanishing of
covariant virtual work rather than by an a priori extremality condition on an
action functional.

The covariant virtual work structure thus accommodates both
integrable (variational) and non-integrable tensorial contributions.

\textit{Covariant d’Alembert-Lagrange principle.—}
The covariant d’Alembert-Lagrange principle distinguishes two logically
independent levels. At the \emph{Lagrangian} level, the first variation of the
action admits an intrinsic decomposition into bulk and boundary terms under
arbitrary metric variations. This decomposition is purely variational and does
not by itself impose mechanical equilibrium.

Dynamical content enters only at the \emph{d’Alembertian} level, where the
variation is restricted to the admissible class of virtual displacements and
equilibrium is imposed through the vanishing of the corresponding total virtual work.

Constraints may be incorporated at the Lagrangian level through auxiliary
Lagrange multipliers
\cite{anderson1971cosmological,BUCHMULLER1988292,CAPOZZIELLO2010198,GAO2011107,Kamenshchik2020,kichenassamy1986lagrange,Safko1976Lagrange}.
We therefore consider an action of the form
\begin{equation}
S_\lambda
=
\int_V d^4x\, \sqrt{-g}\,\lambda\, f ,
\label{lambd}
\end{equation}
where $\lambda$ is an undetermined Lagrange multiplier and
$f(g_{\mu\nu},\partial_\sigma g_{\mu\nu})$
denotes a constraint function. Such constraints may be local,
acting pointwise on the fields, or global, arising from integral
conditions on the configuration. A prototypical example is provided by
isoperimetric constraints of the form
$
\int_V \sqrt{-g}\,F(g_{\mu\nu},\partial_\sigma g_{\mu\nu})=\mathrm{constant},
$
for which the associated multiplier is a constant, fixed by the global
constraint rather than by local equations
\cite{lanczos1970variational,CourantHilbert1953,GelfandFomin1963}.
This distinction reflects a structural property of the constraint: global
conditions impose a single restriction on the configuration and therefore
introduce constant multipliers, whereas local constraints generically lead
to spacetime-dependent multiplier fields.

The total gravitational action on a spacetime region $V$ with boundary
$\partial V$ then takes the form
\begin{equation}
S_{GR}
=
\int_V d^4x\,\sqrt{-g}
\left(
\frac{1}{2\kappa}R
+
\mathcal{L}_M
+
\lambda f
\right).
\end{equation}

The EH term contains second derivatives of the metric, so its first variation yields a second–order bulk operator together with boundary contributions. To maintain a consistent differential structure, the matter and constraint sectors may be taken to depend on the metric and its first derivatives (see Appendix~\ref{appendix:lagrangian}), in agreement with the general definition of the stress–energy tensor \cite{gron2007einstein}.

The bulk metric variations of the gravitational, matter, and constraint
sectors combine to give
\begin{equation}
\delta S_{\mathrm{bulk}}
=
\frac12
\int_V d^4x\,\sqrt{-g}\,
\left(
\frac{1}{\kappa} G_{\mu\nu}
-
T_{\mu\nu}
+
C_{\mu\nu}
\right)
\delta g^{\mu\nu},
\label{eq:DAlembertBulk}
\end{equation}
with $\kappa = 8\pi G$, where
\begin{equation}
T_{\mu\nu} =
-\frac{2}{\sqrt{-g}}
\left(
\frac{\partial (\sqrt{-g}\mathcal{L}_M)}{\partial g^{\mu\nu}}
-
\partial_\sigma
\frac{\partial (\sqrt{-g}\mathcal{L}_M)}
{\partial(\partial_\sigma g^{\mu\nu})}
\right)
\end{equation}
is the energy–momentum tensor, and
\begin{equation}
C_{\mu\nu}
=
-\,\frac{2}{\sqrt{-g}}
\left(
\frac{\partial(\sqrt{-g}\lambda f)}{\partial g^{\mu\nu}}
-
\partial_\sigma
\frac{\partial(\sqrt{-g}\lambda f)}
{\partial(\partial_\sigma g^{\mu\nu})}
\right)
\end{equation}
is the reaction tensor associated with the constraint sector. Any dependence on additional fields such as a scalar $\psi$ is
implicitly included in the definitions of both $T_{\mu\nu}$ and
$C_{\mu\nu}$ through the corresponding Lagrangian densities.

The full first variation therefore decomposes as
\begin{equation}
\delta S_{total}
=
\delta S_{\mathrm{bulk}}
+
\delta S_{\partial V},
\end{equation}
where $\delta S_{\partial V}$ collects the surface terms arising from the
integrations by parts.

Equation~\eqref{eq:DAlembertBulk} identifies the symmetric tensor
\begin{equation}
A_{\mu\nu}
\equiv
G_{\mu\nu}
-
\kappa T_{\mu\nu}
+
\kappa C_{\mu\nu}
\end{equation}
that appears in the bulk variational identity. At this stage no field equation
has been imposed: $A_{\mu\nu}$ is defined \emph{off–shell} as the residual tensor
field of the action.

Hamilton’s principle formulates dynamics through action extremization under arbitrary variations with prescribed boundary terms. By contrast, the d’Alembert–Lagrange principle expresses dynamics as an equilibrium condition on admissible virtual displacements without imposing prior assumptions on the boundary terms. At this stage, this distinction naturally leads to the selection of the latter as the more general framework for formulating the dynamics. 

Therefore we introduce dynamics by imposing an equilibrium condition on the
\emph{total} covariant virtual work,
identified with the first variation of the total action,
\begin{equation}
\delta W_{\mathrm{total}}
\equiv
\delta S_{\mathrm{total}},
\end{equation}
evaluated on admissible virtual displacements
$\delta g^{\mu\nu}_{\mathrm{adm}}$
\cite{Gantmakher1970,lanczos1970variational}.

The covariant d’Alembert–Lagrange principle is then stated as
\begin{equation}
\delta W_{\mathrm{total}}
=
\delta W_{\mathrm{bulk}}
+
\delta W_{\mathrm{boundary}}
=
0 ,
\label{eq:dalembert_total}
\end{equation}
for all admissible virtual displacements
$\delta g^{\mu\nu}_{\mathrm{adm}}$.
In mechanical terms, the integrand represents the dual pairing defining the
total covariant virtual work between the generalized force tensor and the
admissible virtual displacements. 
In the present integrable realization of
the framework, this generalized force tensor is represented by the
variational residual tensor \(A_{\mu\nu}\).

For virtual displacements with compact support in the interior of $V$
\cite{Lee2013}, the boundary contribution vanishes, and equilibrium reduces to
\begin{equation}
\int_V d^4x\,\sqrt{-g}\,
A_{\mu\nu}\,
\delta g^{\mu\nu}_{\mathrm{adm}}
=
0 .
\end{equation}

Since admissible compactly supported variations are locally arbitrary within
the admissible sector of configuration space, the vanishing of this pairing
implies the local equilibrium equation
\begin{equation}
A_{\mu\nu}
\equiv
G_{\mu\nu}
-
\kappa T_{\mu\nu}
+
\kappa C_{\mu\nu}
=
0 .
\end{equation}

This condition defines the covariant d’Alembert–Lagrange formulation of GR as the realization of the principle in the gravitational dynamics.

We call \emph{on-shell} those configurations $g_{\mu\nu}$ for which
$A_{\mu\nu}=0$ pointwise in $V$. Although the equilibrium condition is
formulated in configuration space, it defines a local tensorial identity in
spacetime. This condition represents the covariant equilibrium of dynamical
contributions in the covarant virtual work formulation. In this sense, curvature provides the covariant realization of inertia:
$G_{\mu\nu}$ encodes the intrinsic inertial response of spacetime geometry,
while $T_{\mu\nu}$ represents stress--energy and $C_{\mu\nu}$ the
constraint-induced reaction.  

In vacuum, the equilibrium condition reduces to $G_{\mu\nu}=0$, corresponding to configurations in which equilibrium is satisfied purely through the geometric sector.

The derivation imposes no restriction on the internal structure of
$T_{\mu\nu}$ beyond symmetry and regularity, reflecting the fact that the
equilibrium condition is formulated independently of any constitutive
assumptions \cite{Papastavridis2002}. In particular, it accommodates anisotropic, viscous, and
dissipative stresses, in agreement with classical continuum formulations and
with modern extensions of the d’Alembert--Lagrange framework to dissipative
systems, including settings relevant to nonequilibrium thermodynamics
\cite{Papastavridis2002,GayBalmazYoshimura2017b,GayBalmazYoshimura2017c}.
This generality is a direct consequence of the formulation: while the geometric
sector, through $G_{\mu\nu}$, encodes the inertial response required by the
equilibrium condition, the stress--energy tensor remains unrestricted at the
structural level. Accordingly, the principle constrains only the generalized force
tensor $A_{\mu\nu}$ through the total covariant virtual work condition, leaving the
stress-energy sector free of any constitutive ansatz, and remains compatible
with the general Hawking-Ellis classification
\cite{Segre1884Omografie,osti_4656212,Hawking:1973uf,MartinMoruno:2021niw}.

Boundary contributions are treated as an intrinsic part of the covariant
mechanical equilibrium. Their structure encodes the preceding analysis and
fully characterizes the boundary sector without the need to impose
external prescriptions. In this sense, while related to standard boundary
formulations \cite{York:1986lje,Gibbons:1976ue,GibbonsHawking1977,Parattu:2015gga,Chakraborty:2016yna, Lee:1990nz, Iyer:1994ys, Wald:1993nt, Donnelly:2016auv, Jubb:2016qzt, Hopfmuller:2016scf},
they are incorporated here directly within the d’Alembert–Lagrange framework.

Accordingly, the admissible set of virtual displacements has a hierarchical
structure: compactly supported variations determine local bulk equilibrium,
whereas boundary-supported variations probe exchange and admissibility at
$\partial V$, without reintroducing bulk dynamics.

If the admissible class of virtual displacements is reversible, equilibrium is
expressed by Eq.~\eqref{eq:dalembert_total}. When admissibility becomes
one-sided, as occurs at the boundary of configuration space
\cite{lanczos1970variational,Rumyantsev,LeineAeberhardGlocker2009},
the equilibrium postulate generalizes to the inequality
\begin{equation}
\delta_X W_{\mathrm{bulk}}+\delta_X W_{\mathrm{boundary}}\le 0 ,
\label{eq:FourierIneq}
\end{equation}
with the sign determined by orientation. Equation~\eqref{eq:FourierIneq}
constitutes the covariant analogue of Fourier’s inequality in analytical
mechanics and captures equilibrium under directional admissibility.

This structure differs from the Hamilton’s principle. While fixed
boundary data lead to equality conditions, the d’Alembert–Lagrange framework
allows the admissible set itself to encode reversibility or unilateral
admissibility, so
that equilibrium arises either as an equality or as an inequality. Although
related extensions of Hamilton’s principle exist for unilateral constraints
\cite{LeineAeberhardGlocker2009,Rumyantsev}, here the inequality structure follows directly from the restriction of admissible virtual displacements rather than from a modified action principle. 

Having fixed local bulk equilibrium through admissible compactly supported
virtual displacements, we next implement a global isoperimetric constraint on the
spacetime four-volume to generate an effective cosmological term as an 
geometric reaction.

\textit{Isoperimetric constraint and cosmological constant.—} Global admissibility conditions impose integral restrictions on configuration
space rather than pointwise constraints on the fields. Isoperimetric conditions
\cite{CourantHilbert1953,lanczos1970variational,GelfandFomin1963} fix the
spacetime four–volume,
\begin{equation}
\Phi[g] \equiv \int_V d^4x\,\sqrt{-g} = V_0 ,
\end{equation}
and are implemented through the Lagrange multiplier term
\begin{equation}
S_\lambda
=
-\frac{\lambda}{2\kappa}\,\Phi[g].
\end{equation}
Because the constraint is global, it imposes a single restriction on the
configuration and therefore introduces a constant multiplier, in contrast with
local constraints which generically yield spacetime-dependent multipliers.

At the Lagrangian level, variation of $S_\lambda$ produces a reaction tensor
proportional to the metric. The bulk variational source entering the
d’Alembert–Lagrange equilibrium becomes
\begin{equation}
A_{\mu\nu}
\equiv
G_{\mu\nu}
-
\kappa T_{\mu\nu}
+
\Lambda g_{\mu\nu},
\qquad
\Lambda \equiv \lambda/2 .
\end{equation}

Although the constraint is global, equilibrium remains local: compactly
supported admissible variations enforce the vanishing of bulk virtual work
pointwise, yielding
\begin{equation}
G_{\mu\nu}+\Lambda g_{\mu\nu}=\kappa T_{\mu\nu}.
\end{equation}
The multiplier $\Lambda$ enters as a globally fixed reaction parameter. Taking
the trace and integrating over $V$ gives
\begin{equation}
\Lambda=\frac{1}{4V_0}\int_V d^4x\,\sqrt{-g}\,(R+\kappa T),
\end{equation}
showing that $\Lambda$ is not an independent coupling but is determined by the
global trace content of the admissible configuration, with constancy following
from the global constraint. The cosmological term thus arises as a structural
consequence of global admissibility conditions in configuration space.

From the covariant d’Alembert–Lagrange principle, the cosmological term arises as an \emph{ideal} reaction associated with a global constraint. By Postulate~A, it performs no virtual work on admissible displacements $\delta g^{\mu\nu}_{\mathrm{adm}}$ and therefore modifies the equilibrium equations without altering the virtual work equilibrium (see Appendix~\ref{appendix:isoperimetric_derivation}).

This structural origin identifies the cosmological term as a global geometric reaction controlled by spacetime averages.

The appearance of spacetime averages connects this framework with
approaches in which global variables enter the gravitational field equations,
such as vacuum-energy sequestering models and related constructions
\cite{KaloperPadilla2014,CarrollRemmen2017,ArkaniHamed2002,Woodard2014}.
Within the present formulation, such structures arise as direct consequences
of admissibility conditions, indicating that the nonlocal character of the
cosmological term reflects a fundamental restriction on admissible
gravitational configurations.

This mechanism differs from the standard formulation of GR, where $\Lambda$ is
introduced as a fundamental coupling in the action and enters the field
equations kinematically \cite{einstein1917cosmological}. It also differs from
unimodular gravity
\cite{anderson1971cosmological,BUCHMULLER1988292,Unruh1989,HenneauxTeitelboim1989},
where a local constraint yields the traceless Einstein equations and $\Lambda$
appears as an integration constant. In contrast, the present constraint acts
on configuration space while leaving the trace sector dynamical, so that the
full Einstein equations follow directly from the covariant d’Alembert–Lagrange equilibrium condition, with $\Lambda$ fixed by a global
constraint through spacetime averages rather than by integration freedom.

This interpretation has direct physical implications: the value of $\Lambda$
is set by spacetime averages of the dynamical fields rather than by microscopic
vacuum contributions, suggesting that its smallness may reflect global
properties of the spacetime configuration rather than a fine-tuned
cancellation of vacuum energy.

\textit{Conclusions and outlook.—} We have established a covariant virtual work structure  for spacetime geometry together with a mechanical admissibility criterion encoded
in Postulate~A, and on this basis formulated a covariant d’Alembert–Lagrange principle for spacetime geometry. Gravitational dynamics
is expressed as the vanishing of total covariant virtual work, and the
EFE arise as local equilibrium relations rather than as consequences of an extremal action principle.

The construction further establishes a mechanical classification of gravitational reactions through Postulate~A: ideal contributions do not enter the total covariant virtual work, whereas non-ideal sectors contribute explicitly to it and therefore enter directly into the covariant d’Alembert–Lagrange principle; both sectors nevertheless modify the resulting EFE through the equilibrium conditions.  This defines a general mechanical structure for constraint-induced contributions within GR without introducing additional curvature invariants or ad hoc modifications of the gravitational action. Admissibility may be reversible or one-sided, so that the equilibrium condition extends from equality to inequality at the boundary of configuration space, yielding a covariant inequality structure with no counterpart in the Hamiltonian variational formulation of GR.

The Einstein tensor $G_{\mu\nu}$ acquires a direct mechanical interpretation: it encodes the intrinsic inertial response of spacetime geometry, while $T_{\mu\nu}$ represents stress--energy and constraint-induced contributions, encoded in the reaction tensor $C_{\mu\nu}$. The formulation imposes no restriction on the internal structure of $T_{\mu\nu}$ beyond symmetry and regularity, accommodating general stress tensors,
including anisotropic and dissipative sectors. The equilibrium condition thus places inertial and gravitational contributions on the same footing within a
single covariant equilibrium condition. In the classical limit, where the
metric is non-dynamical, this geometric response becomes effectively fixed and inertial behavior reduces to motion on a prescribed background. In this way, inertia is identified as a geometric response at the level of the
equilibrium principle, with its apparent distinction from gravitation
emerging only in appropriate limits.

A central structural result is that global admissibility conditions generate
geometric reaction contribution. In particular, a global
isoperimetric constraint on spacetime four–volume produces a cosmological
term as an ideal reaction, with its value fixed by the global trace content
of the admissible configuration. The resulting cosmological constant is therefore neither a fundamental coupling nor an integration constant, but instead a global reaction parameter determined by admissibility.

Within this mechanical framework, the appearance of spacetime averages in the
reaction sector connects this construction with related nonlocal approaches.
Here, such structures arise as consequences of admissibility conditions,
indicating that their nonlocal character reflects a fundamental restriction on
admissible gravitational configurations. The isoperimetric constraint provides
the simplest realization of this structure. 

As a consequence, the cosmological term need not be associated with a local
vacuum-energy density, suggesting that observational interpretations based on
vacuum energy may warrant reconsideration.

Within this framework, theories in which constraints are implemented through
Lagrange multipliers admit a mechanical interpretation in terms of reaction
tensors. Under Postulate~A, this formulation provides a consistent setting for
such constrained sectors within gravitational dynamics. In particular, when these contributions arise as ideal reactions, as in the isoperimetric construction presented here, the resulting field equations remain consistent with Lovelock’s theorem: no additional curvature invariants are introduced, and the equations retain their second-order character, with these terms appearing as reactions rather than as modifications of the gravitational Lagrangian.

A natural extension of this framework concerns the relation with the
equivalence principle. The identification of geometric response with inertia
suggests that the equivalence between inertial and gravitational effects may
admit a structural interpretation within this formulation.

Further developments involve the analysis of boundary contributions and of
inequality constraints arising from one-sided admissibility conditions. Such
inequalities define a geometric route toward entropy-like structures, in
connection with variational formulations of nonequilibrium thermodynamics, and
suggest the possibility of intrinsic inequality principles at the level of
configuration space. A fully covariant identification of entropy within this
framework remains to be established.

In this framework, gravitational dynamics is formulated as a covariant
equilibrium structure in configuration space, in which geometric response,
matter sources, and constraint-induced reaction sectors are unified through
admissibility conditions. By formulating the dynamics through the covariant
d'Alembert--Lagrange principle rather than solely through action
extremization, the Einstein field equations acquire a structurally broader
mechanical framework. This perspective accommodates dissipative and non-ideal sectors
beyond purely conservative settings. Boundary contributions can likewise
remain within the covariant d'Alembert--Lagrange framework, indicating that
their analysis may reveal additional geometric and mechanical structures.
Within this structure, the cosmological constant emerges as a global reaction
parameter fixed by admissibility, reflecting an intrinsically nonlocal
geometric origin. 


\textit{Acknowledgments.—}
A.C. acknowledges J.~Chagoya, A.~Roque, and R.~Hernández for helpful discussions.
This work was supported by SECIHTI grant DCF-320821.
A.C. acknowledges support from CONAHCyT grant 847964.

\appendix

\renewcommand{\thesection}{\Alph{section}}
\renewcommand{\theequation}{\thesection\arabic{equation}}
\setcounter{section}{0}
\refstepcounter{section}
\section*{APPENDIX \thesection: General metric variation}
\label{appendix:lagrangian}
\setcounter{equation}{0}

We consider a Lagrangian density depending on the metric and its first
partial derivatives,
\begin{equation}
\mathfrak{L}
=
\mathfrak{L}
\!\left(
g^{\mu\nu},
\partial_\sigma g^{\mu\nu}
\right),
\qquad
S=\int_V d^4x\,\mathfrak{L}.
\end{equation}

The metric variation reads
\begin{align}
\delta\mathfrak{L}
&=
\frac{\partial\mathfrak{L}}{\partial g^{\mu\nu}}
\,\delta g^{\mu\nu}
+
\frac{\partial\mathfrak{L}}{\partial(\partial_\sigma g^{\mu\nu})}
\,\partial_\sigma \delta g^{\mu\nu}.
\end{align}

Integrating by parts yields
\begin{align}
\delta\mathfrak{L}
=
\mathcal{F}_{\mu\nu}(\mathfrak{L})\,\delta g^{\mu\nu}
+
\partial_\sigma \mathcal{H}^\sigma(\mathfrak{L}),
\end{align}
where
\begin{align}
\mathcal{F}_{\mu\nu}(\mathfrak{L})
&=
\frac{\partial\mathfrak{L}}{\partial g^{\mu\nu}}
-
\partial_\sigma
\left(
\frac{\partial\mathfrak{L}}{\partial(\partial_\sigma g^{\mu\nu})}
\right),
\\
\mathcal{H}^\sigma(\mathfrak{L})
&=
\frac{\partial\mathfrak{L}}{\partial(\partial_\sigma g^{\mu\nu})}
\,\delta g^{\mu\nu}.
\end{align}

Consequently,
\begin{align}
\delta S
=
\int_V d^4x\,
\mathcal{F}_{\mu\nu}(\mathfrak{L})\,\delta g^{\mu\nu}
+
\int_{\partial V} d\Sigma_\sigma\,
\mathcal{H}^\sigma(\mathfrak{L}).
\end{align}

This structure applies in particular to the matter and constraint sectors
considered in the main text.

\refstepcounter{section}
\section*{APPENDIX \thesection: Isoperimetric constraint and the ideal reaction tensor}
\label{appendix:isoperimetric_derivation}
\setcounter{equation}{0}

We collect here the variational steps associated with the global
isoperimetric constraint introduced in the main text,
\begin{equation}
\Phi[g] \equiv \int_V d^4x\,\sqrt{-g}= V_0 .
\end{equation}
This defines a global restriction on the metric configuration space without
imposing a pointwise condition on the metric.

The constraint is incorporated through the Lagrange multiplier term
\begin{equation}
S_\lambda = -\frac{\lambda}{2\kappa}\Phi[g].
\end{equation}

Because the restriction fixes a single global functional rather than a
field of local constraints, the associated multiplier is a constant parameter,
not a spacetime-dependent field.

Under an arbitrary metric variation,
\begin{equation}
\delta \sqrt{-g}
=
-\frac{1}{2}\sqrt{-g}\, g_{\mu\nu}\,\delta g^{\mu\nu},
\end{equation}
so that
\begin{equation}
\delta \Phi[g]
=
-\frac{1}{2}
\int_V d^4x\,\sqrt{-g}\,
g_{\mu\nu}\,\delta g^{\mu\nu}.
\end{equation}

The variation of the constraint term becomes
\begin{equation}
\delta S_\lambda
=
\frac{1}{2\kappa}
\int_V d^4x\,\sqrt{-g}\,
\left(
\frac{\lambda}{2}g_{\mu\nu}
\right)
\delta g^{\mu\nu}.
\end{equation}

Comparing with
\begin{equation}
\delta S_\lambda
=
\frac{1}{2\kappa}
\int_V d^4x\,\sqrt{-g}\,
C_{\mu\nu}\delta g^{\mu\nu},
\end{equation}
one obtains
\begin{equation}
C_{\mu\nu}
=
\Lambda g_{\mu\nu},
\qquad
\Lambda=\lambda/2 .
\end{equation}

The admissible variations are obtained by linearizing the constraint
\[
\Phi[g]=V_0 .
\]
Let
\[
g^{\mu\nu}(\epsilon)
=
g^{\mu\nu}
+
\epsilon\,\delta g^{\mu\nu}_{\mathrm{adm}}
+
O(\epsilon^2)
\]
be a curve tangent to the constrained configuration subspace at
\(\epsilon=0\). Since the background configuration already satisfies
\(\Phi[g]=\Phi_0\), admissibility requires preservation of the constraint to
first order,
\[
\Phi[g(\epsilon)]
=
V_0
+
\epsilon\,\delta\Phi[g]
+
O(\epsilon^2).
\]
Compatibility with the constraint therefore requires the linear term to
vanish. Hence,
\begin{equation}
\delta\Phi[g]
=
\left.
\frac{d}{d\epsilon}
\Phi[g(\epsilon)]
\right|_{\epsilon=0}
=
0.
\end{equation}

Using the previously obtained variation of \(\Phi[g]\), one finds
\begin{equation}
\delta\Phi[g]
=
-\frac{1}{2}
\int_V d^4x\,\sqrt{-g}\,
g_{\mu\nu}\delta g^{\mu\nu}_{\mathrm{adm}},
\end{equation}
and therefore
\begin{equation}
\int_V d^4x\,\sqrt{-g}\,
g_{\mu\nu}\delta g^{\mu\nu}_{\mathrm{adm}}
=
0.
\label{eq:isoperimetric_admissible}
\end{equation}

This equation characterizes the admissible tangent space associated with the
global isoperimetric constraint. The condition is global rather than
pointwise: admissible variations are not required to satisfy
\[
g_{\mu\nu}\delta g^{\mu\nu}_{\mathrm{adm}}=0
\]
locally, but only the integrated restriction imposed by the fixed
four--volume constraint.

Since
$\displaystyle \frac{\delta\Phi[g]}{\delta g^{\mu\nu}}
=
-\frac{1}{2}\sqrt{-g}\,g_{\mu\nu}$,
and
$C_{\mu\nu}=\Lambda g_{\mu\nu}$,
the reaction tensor is proportional to the functional gradient of the
isoperimetric constraint. Hence,
\[
C_{\mu\nu}\in \mathrm{Ann}(T\mathcal C_{\mathrm{adm}}),
\]
so that the reaction acts along the normal direction to the admissible
configuration subspace.

Consequently,
\begin{equation}
\int_V d^4x\,\sqrt{-g}\,
C_{\mu\nu}\delta g^{\mu\nu}_{\mathrm{adm}}
=
\Lambda
\int_V d^4x\,\sqrt{-g}\,
g_{\mu\nu}\delta g^{\mu\nu}_{\mathrm{adm}}
=0,
\end{equation}
where the last equality follows directly from
\eqref{eq:isoperimetric_admissible}. The isoperimetric reaction therefore
performs no virtual work on admissible virtual displacements.

The vanishing of virtual work does not follow solely from the holonomic
character of the constraint, but from the specific form of the reaction
tensor. Although the admissibility condition is global, the associated
reaction tensor acts locally through the pointwise virtual work pairing
\(C_{\mu\nu}\delta g^{\mu\nu}\).

The isoperimetric constraint is therefore classified as ideal in the sense of
Postulate~A.

\bibliographystyle{apsrev4-2}
\bibliography{bibliography}


\end{document}